\documentclass[12pt,aps,pre,notitlepage,superscriptaddress,floatfix]{revtex4-1}
\usepackage{graphicx}
\usepackage{amssymb,amsmath,amsbsy,amsfonts}
\usepackage{epsfig}
\usepackage[percent]{overpic}
\usepackage{color}

\begin{document}

\title{Formation and growth of shear bands in glasses: existence of 
an underlying directed percolation transition}

\author{Gaurav Prakash Shrivastav}
\affiliation{Institut f\"ur Theoretische Physik II, Heinrich-Heine-Universit\"at 
D\"usseldorf, Universit\"atsstr. 1, 40225 D\"usseldorf, Germany}
\author{Pinaki Chaudhuri}
\affiliation{The Institute of Mathematical Sciences, CIT Campus, Taramani, 
Chennai 600 113, India}
\author {J\"urgen Horbach}
\affiliation{Institut f\"ur Theoretische Physik II, Heinrich-Heine-Universit\"at 
D\"usseldorf, Universit\"atsstr. 1, 40225 D\"usseldorf, Germany}

%%%%%%%%%%%%%%%%%%%%%%%%%%%%%%%%%%%%%%%%%%%%%%%%%%%%%%%%%%%%%%%%%%%%%%%%%%%%%%%%%%
%
\begin{abstract}
The response of glasses to mechanical loading often leads to the formation
of inhomogeneous flow patterns.  Among them, shear bands, associated
with strain localization in form of band-like structures, are ubiquitous
in a wide variety of materials, ranging from soft matter systems to metallic alloys.  
They can be precursors to catastrophic
failure, implying that a better understanding of the underlying mechanisms
of shear banding could lead to the design of smarter materials. Here,
molecular dynamics simulations are used to reveal the formation of shear
bands in a binary Lennard-Jones glass, subject to a constant strain
rate. At a critical strain, this system exhibits for all considered
strain rates a transition towards the formation of a percolating cluster
of mobile regions. We give evidence that this transition belongs to the
universality class of directed percolation. Only at low shear rates,
the percolating cluster evolves into a transient (but long-lived) shear
band with a diffusive growth of its width.
\end{abstract}
%
%%%%%%%%%%%%%%%%%%%%%%%%%%%%%%%%%%%%%%%%%%%%%%%%%%%%%%%%%%%%%%%%%%%%%%%%%%%%%%%%%%
%\newpage

\maketitle

{\it Introduction.} An external shear field leads in general to a
rejuvenation of the glass state, transforming the amorphous solid
into a flowing fluid \cite{rodneyrev2011,barratlemaitrerev}.  Under a
constant strain rate, the transition to plastic flow can be located via
the dependence of the shear stress on the applied strain. It is marked
by a maximum in the stress-strain relation which for a simple planar
Couette flow geometry occurs typically at a strain of the order of
0.1 \cite{barratlemaitrerev,zausch08,falk-maloney-2010,knowlton}.
Beyond this maximum, the system evolves into a steady-state regime where
the system displays a homogeneous flow pattern, e.g.~in the case of
planar Couette flow manifested as a linear velocity profile. However,
the strain necessary to reach this steady-state regime depends on many
factors such as the history of the initial undeformed glass state and the
applied strain rate \cite{moorcroft-cates-fielding-11,varnik04}.  In the
transient regime before the steady state is reached, the occurrence
of spatially inhomogeneous flow patterns is very common. Often, such
inhomogeneous response leads to the formation of shear bands, with
co-existing regions of contrasting mobilities spanning large scales
\cite{schuhrev}. Even when the mechanical loading is switched off,
the shear-band regions can be imprinted as frozen-in structures in the
resulting solid  and thus strongly affect its material properties
\cite{wilde11,zhang-wang-greer-06}. Therefore, the microscopic
understanding of the formation and growth of shear bands in glassy solids
is a highly-debated issue in material science.

Despite the fact that shear banding is ubiquitous in amorphous
materials \cite{mb08,schuhrev,fs14,gm13,divouxrev15}, the microscopic
processes that lead to the formation of such complex structures still
remain ill understood.  Unlike the case of micellar systems or granular
materials, where shear bands mainly form in the regime of large applied
shear rates, driven by a coupling between structures and external
shear \cite{sh10,ir14,fs14,mb12}, the scenario in glasses seems to be
different. The inhomogeneous response in the case of amorphous solids
tends to be observed at small shear rates, with the spatio-temporal
extent becoming more prominent as one approaches the yielding threshold
\cite{rodneyrev2011,barratlemaitrerev,chboc12,mb11}.  And, so far,
there is no indication of any underlying macroscopic mechanical
instability driving the formation of shear bands. While there have
been several observations in experiments \cite{vp11,bp10}, numerical
simulations \cite{vb03,sf06,bailey06,ch13,ratul-procaccia-12} and
phenomenological models \cite{moorcroft-fielding-13,damienroux2011},
a quantitative analysis that allows to predict the conditions under
which shear bands form in glasses is lacking.

The question about the origin of shear bands is intimately
related to the question how flow is initiated in an amorphous
solid under applied shear. Recent studies suggest that the
response of the glass to the applied strain is governed by local
heterogeneities that are either already present in the undeformed
solid or form during the initial application of the shear field
\cite{rodneyrev2011,barratlemaitrerev,falk-maloney-2010}. These
heterogenities are associated with ``hot spots'' of higher
mobility that grow while the strain of the system increases
\cite{clement12,sentja15}. For athermal amorphous systems,
evidence has been given that the transition towards plastic
flow is provided by an avalanching of the mobile hot spots
\cite{lemaitrecaroli09,bl07,smarajit-procaccia-10}. Imprints of that have
also been reported in thermal systems \cite{chattoraj2010,hentschel10}.
In fact, such avalanching is well-known in the context of self-organized
criticality \cite{btw88}. A prominent example is the sandpile model
\cite{btw88}. Recently, it has been shown that a driven version of
this model belongs to the directed percolation (DP) universality
class \cite{basu14}. In fact, as put forward by the DP conjecture
\cite{janssen81,grassberger82}, DP universality is believed to be very
robust, applying to a broad class of non-equilibrium phase transitions
\cite{hh00}.

Thus -- although not addressed so far -- it is a natural question whether
the onset of flow in a sheared glass is linked to a DP transition. In
this work, we show that this indeed the case and thereby we elucidate
the microscopic mechanism that leads to the formation and growth of
shear bands in glassy systems.

Using a model glass forming system, we use large-scale molecular
dynamics simulations to study the mechanical response of a quiescent
amorphous solid when an external shear rate is imposed on it.
We track the locations of regions of large mobilities and thereby we
reveal the existence of the DP transition driven in the direction of
applied shear. While the DP transition is seen for all shear-rates,
visible shearbanding only only for small shear rates near the yielding
threshold. We demonstrate that the glass eventually fluidizes by the
long-time diffusive invasion of the shear band into the rest of the
system, with the diffusion timescales dependent on the imposed shear
rates. Thus, we provide a quantitative description of how initial local
mobilities build up to the eventual macroscopic flow of the material.

\begin{figure*}
\begin{tabular}{ccc}
\includegraphics[width=4.5cm]{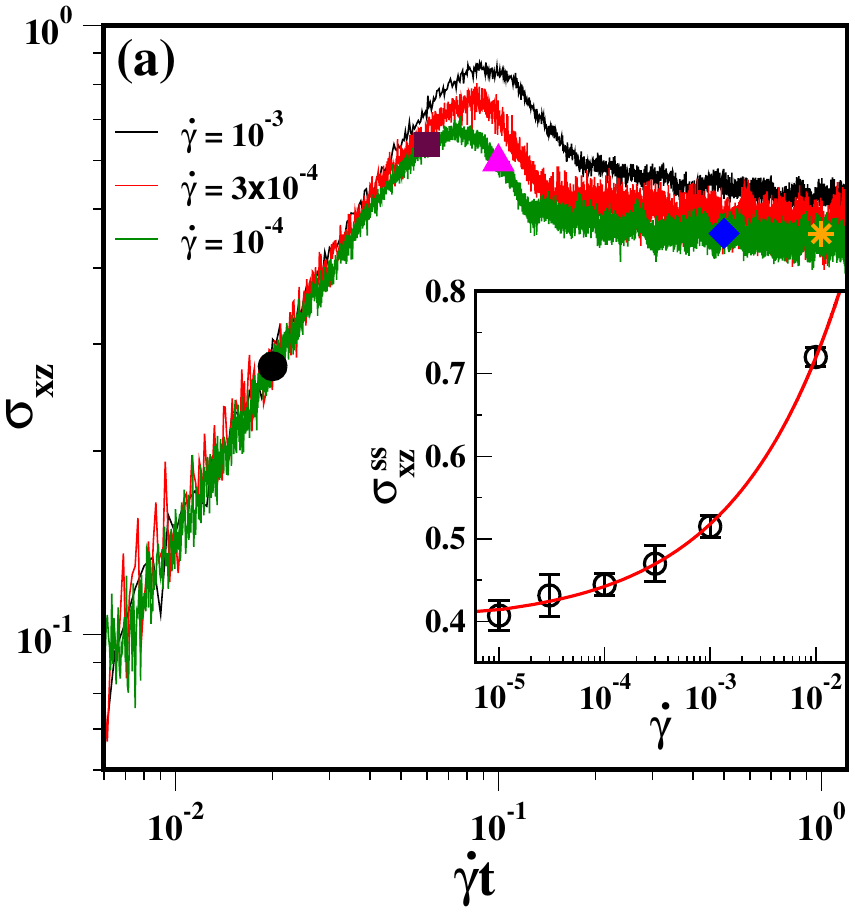}&
  \includegraphics[width=5.0cm]{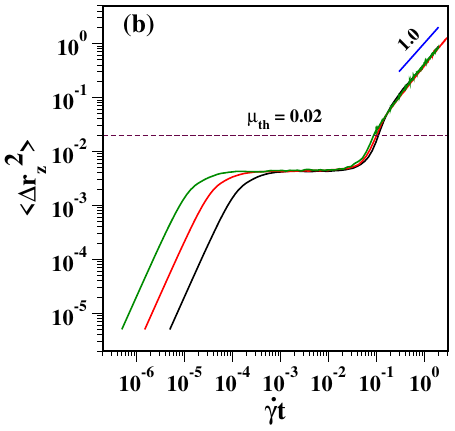}&
  \includegraphics[width=4.8cm]{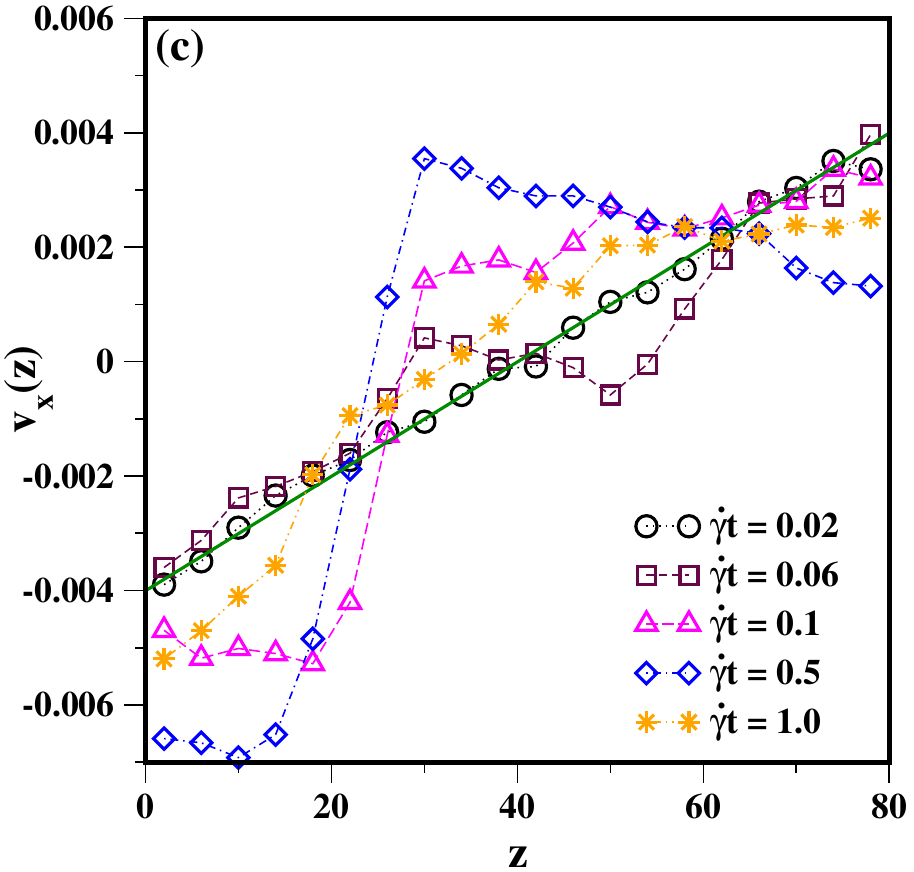}\\
  \includegraphics[width=4.5cm]{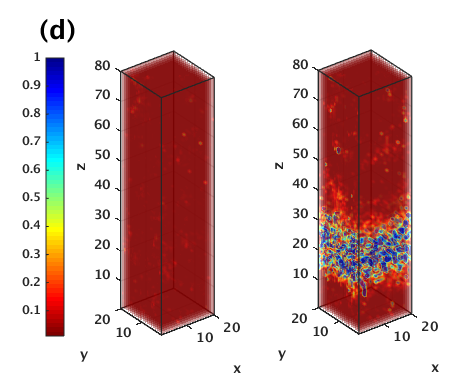}&
  \includegraphics[width=5.0cm]{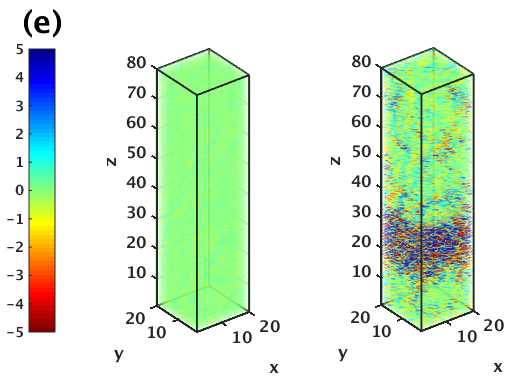}&
  \includegraphics[width=4.1cm]{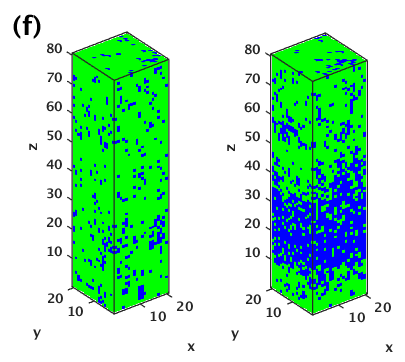}\\
\end{tabular}
\caption{\label{fig1}(a) Stress-strain response of the glass at
temperature $T = 0.2$ for an age of $t_{\textrm{w}}= 10^4$ and sheared
with a constant strain rate.  Data shown for different shear-rates:
$\dot{\gamma} = 10^{-3}, 3\times10^{-4}~\textrm{and} ~10^{-4}$. The
inset shows the flow curve.  Red solid line shows the fitting with
Herschel-Bulkley form $\sigma^{\rm ss}_{xz}(\dot{\gamma}) = 0.3974 +
2.2963{\dot{\gamma}}^{0.43}$.  (b) Variation of $z$-component of the MSD of large particles with
strain. Brown dotted line marks $\mu_{th}( = 0.02)$. (c) Velocity
profiles, for $\dot{\gamma}=10^{-4}$, at five different strain values marked
in (a). Green solid line represents the
expected linear profile. (d) MSD maps at $\dot{\gamma}t = 0.06, 0.5$ for $\dot{\gamma}=10^{-4}$.
(e) Maps of local strain corresponding to MSD maps shown in (d). (f)
Maps of local mobility corresponding to (d). Mobile regions are marked
in blue while immobile regions are marked in green.}
\end{figure*}

{\it Results.} When the shear is applied to the quiescent glass at
$t=0$, the material deforms exhibiting the typical stress ($\sigma$)
vs.~strain ($\dot{\gamma}t$) response [shown in Fig.~\ref{fig1}(a)
for different imposed shear rates $\dot{\gamma}$] with the height of
the overshoot depending on $\dot{\gamma}$ \cite{varnik04}. The measured
steady-state stress as a function of the imposed shear rate is shown in
the inset of Fig.~\ref{fig1}(a); it has the typical Herschel-Bulkley form \cite{varnik04}.
The corresponding single particle dynamics, during this onset of flow,
can be quantified by measuring the non-affine mean-squared displacement (MSD),
$\Delta{r_z}^{2}$, in the direction transverse to the applied shear;
the data is shown in Fig.~\ref{fig1}(b). The particles undergo ballistic
motion at early times and are then caged, before the occurrence of a
super-diffusive regime, prior to diffusion. The onset of super-diffusion
occurs around the stress overshoot in the stress-strain curve, when the
built-up stress is released via the particles breaking their local cages
to subsequently diffuse \cite{zausch08}.

%%%%%%%%%
%%%%%%%%%

In experiments, flow heterogeneities are often diagnosed via the spatial
profiles of local velocities \cite{bp10, divouxprl10}. Similarly, we
measure the spatial profiles of the local flow velocities (averaged
over strain intervals of $0.5\%$), $v_x(z)$, for an imposed shear rate
of $10^{-4}$, at different times after the imposition of shear (marked
on the corresponding stress-strain curve in Fig.~\ref{fig1}(a)). The
spatial profiles are shown in Fig.~\ref{fig1}(c) for one of the initial
states in our ensemble. In the elastic regime, at $\dot{\gamma}t = 0.02$,
the velocity profile is linear, but starts deviating from this shape
as the stress overshoot is approached at $\dot{\gamma}t = 0.06$. This
deviation becomes stronger in the transient regime (at $\dot{\gamma}t
= 0.5$, shown in blue diamonds). Interestingly, the velocity profile
regains its linear shape as plastic flow sets in (at $\dot{\gamma}t =
1.0$, shown in orange stars). The observation of increased heterogeneity,
after the stress-overshoot, is consistent with experimental observations
\cite{divouxprl10}.

However, local velocity profiles only capture the short-time
heterogeneities in dynamics. In order to obtain a more cumulative picture
from $t=0$, we look at the spatially resolved maps of $\Delta{r_z}^{2}(t)$
\cite{ch13, ch14}, by dividing the system into small cells (see Methods
for more details). In Fig.~\ref{fig1}(d), we show the time-evolution
of such a map, for an initial state under the imposed shear rate of
$\dot{\gamma}=10^{-4}$. At a strain of $\dot{\gamma}t = 0.06$, the local
dynamics is nearly homogeneous on this scale. However, at $\dot{\gamma}t
= 0.5$, spatially heterogeneous dynamics is observed, with the more
mobile particles localised in a shear-band-like structure spanning
the $x-y$ plane. We can also construct similar maps of local strain
($\epsilon_{zx}=\partial{\Delta{r_z}}/\partial{x}$), which exhibit a localisation
behaviour similar to the MSD maps; see Fig.~\ref{fig1}(e). Thus,
large local MSDs are also regions of large strains. Henceforth, we use
$\Delta{r_z}^2$ to analyse local dynamical properties.

%%%%%%%%%%%%%%%%%%%%%%%%%%%%%

\begin{figure*}
\begin{center}
\begin{tabular}{cc}
\includegraphics[width=5cm]{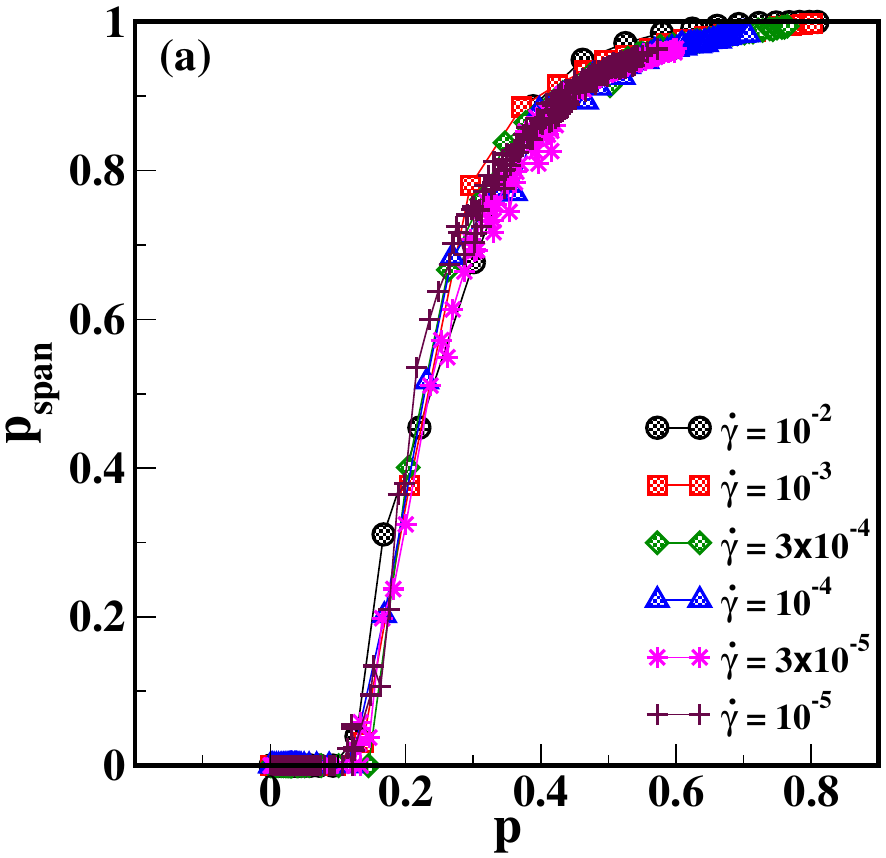}&
   \includegraphics[width=8cm]{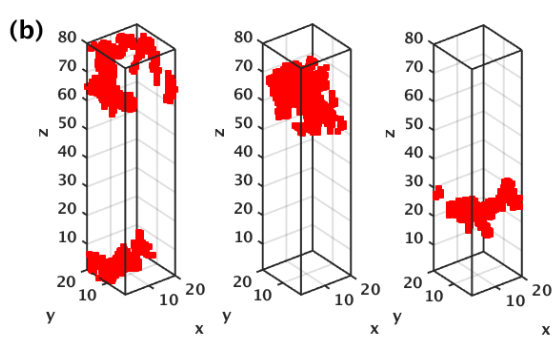}\\
   \includegraphics[width=5cm]{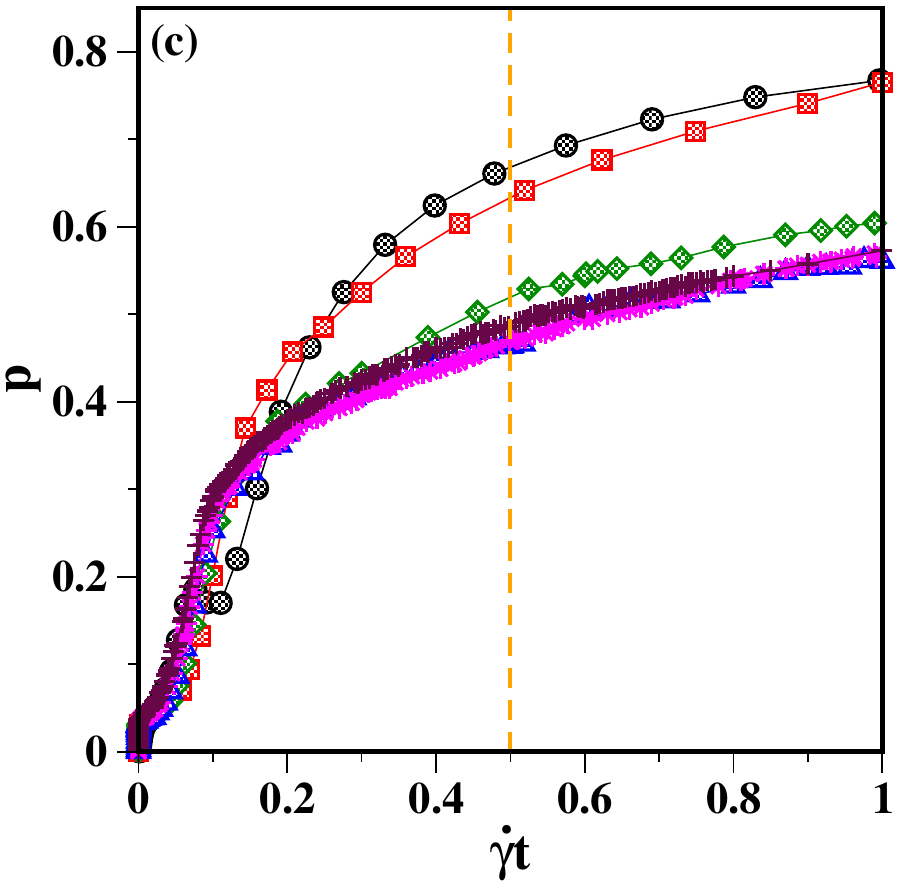}&
   \includegraphics[width=7.5cm]{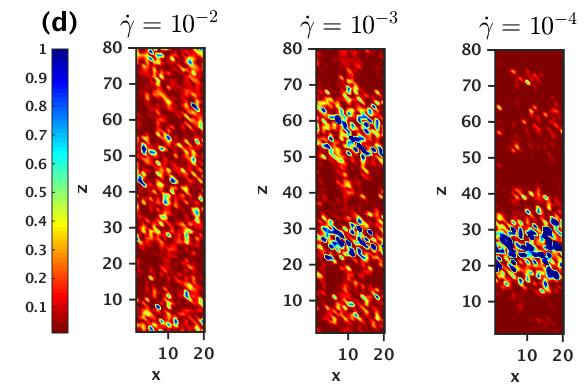}\\
\end{tabular}
\caption{\label{fig3} Occurrence of percolation transition.
(a) Variation of $p_{\textrm{span}}$ with $p$, in the box of dimension $20\times20\times80$,
for strain rates $\dot{\gamma} = 10^{-2},~ 10^{-3},~ 3\times10^{-4},~
10^{-4},~ 3\times10^{-5}~\textrm{and} ~10^{-5}$. (b) Percolating cluster
at the critical point for the different strain rates: $\dot{\gamma} =
10^{-2},~10^{-3}~ \textrm{and}~10^{-4}$ (left to right).  (c) Variation
of $p$ with strain for all strain rates shown in (a). Orange dotted
line corresponds to the $\dot{\gamma}t = 0.5$. (d) 2-Dimensional slice
of MSD map for trajectories shown in (b), at $\dot{\gamma}t = 0.5$.}
\end{center}
\end{figure*}

%%%%%%%%%%%%%%%%%%%%%%%%%%%%%

In order to quantify and characterise the spatio-temporal evolution of
the mobile regions, we define a region to be mobile or not, by setting a
threshold $\mu_{th}=0.02$ on the local $\Delta{r^2}_z$. As marked by the
dashed line in Fig.~\ref{fig1}(c), such a choice of $\mu_{th}$ is larger
than the plateau value in the MSD and thus corresponds to motions beyond
cage-breaking \cite{note}. We then define the local mobility $\psi$ as
\begin{eqnarray}
\label{orderparam}
\psi = \begin{cases} 1 &\mbox{if } \mu \le \mu_{th}, \\
                     0 & \mbox{otherwise},\end{cases}
\end{eqnarray}
where $\mu$ is the average MSD of particles in a sub-box. Following
this convention, we digitize the whole system into mobile and immobile
regions. The mobility maps corresponding to Fig.~\ref{fig1}(d) are shown
in Fig.~\ref{fig1}(f).

%%%%%%%%%%%%%%%%%%%

We now demonstrate that a percolation transition occurs with increasing
strain, involving these mobile regions. As the system evolves under
the applied shear rate, we monitor the fraction of mobile cells, $p$, at any
given instant. Figure~\ref{fig1}(d) suggests that such mobile regions do
form clusters. Thus, we compute what fraction of these mobile regions,
$p_{\rm span}$, is part of a cluster that spans the system, recalling that
such an observable is the order parameter for determining the occurrence
of a percolation transition. In Fig.~\ref{fig3}(a), we plot $p_{\rm
span}$ as a function of $p$, which shows that beyond a critical fraction
$p_c$, all the mobile cells are part of such a spanning cluster. This
indicates the occurrence of a percolation transition of these mobile
cells. Furthermore, we observe that the variation of $p_c$ is nearly
independent of the imposed shear-rate, as seen in Fig.~\ref{fig3}(a)
for a wide range of $\dot{\gamma}$. Thus, the percolation process is
generic to the system's response under shear. In Fig.~\ref{fig3}(b),
we visualise the spanning clusters for three different shear rates, at
the $p_c$ corresponding to each $\dot{\gamma}$, starting from the same
initial state.  Here, we note that the spanning across the system-size
occurs in the direction of flow, which we discuss further later.

%%%%%%%%%%%%%%%

\begin{figure*}
\centerline{
\includegraphics[width=4.6cm]{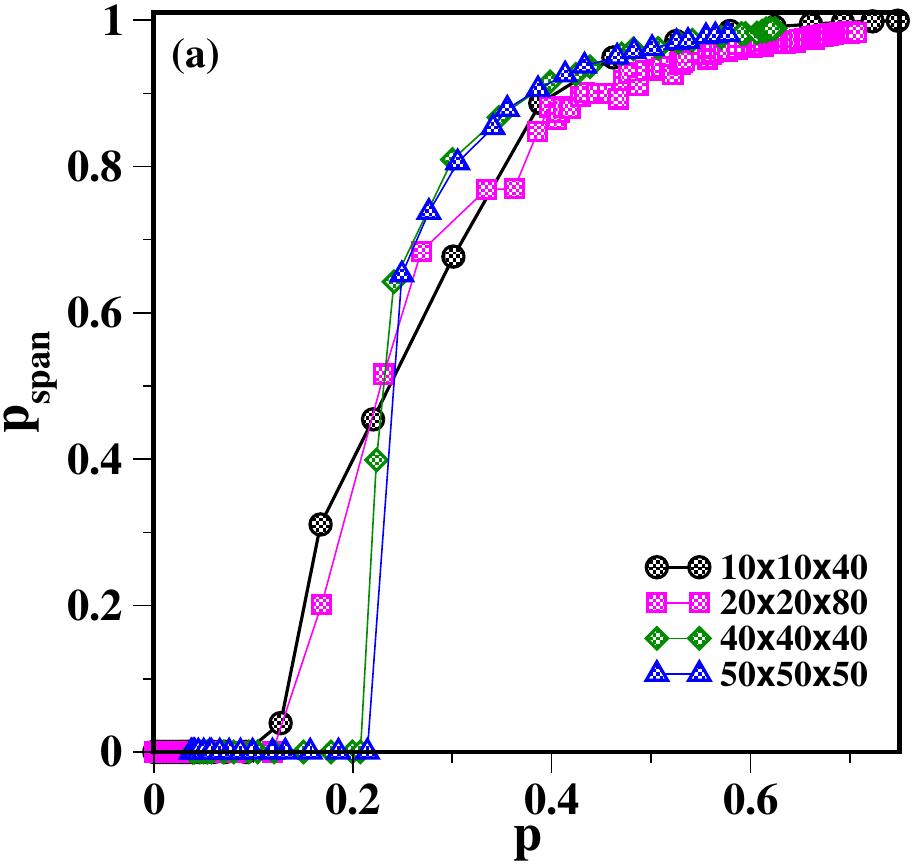}
\includegraphics[width=4.5cm]{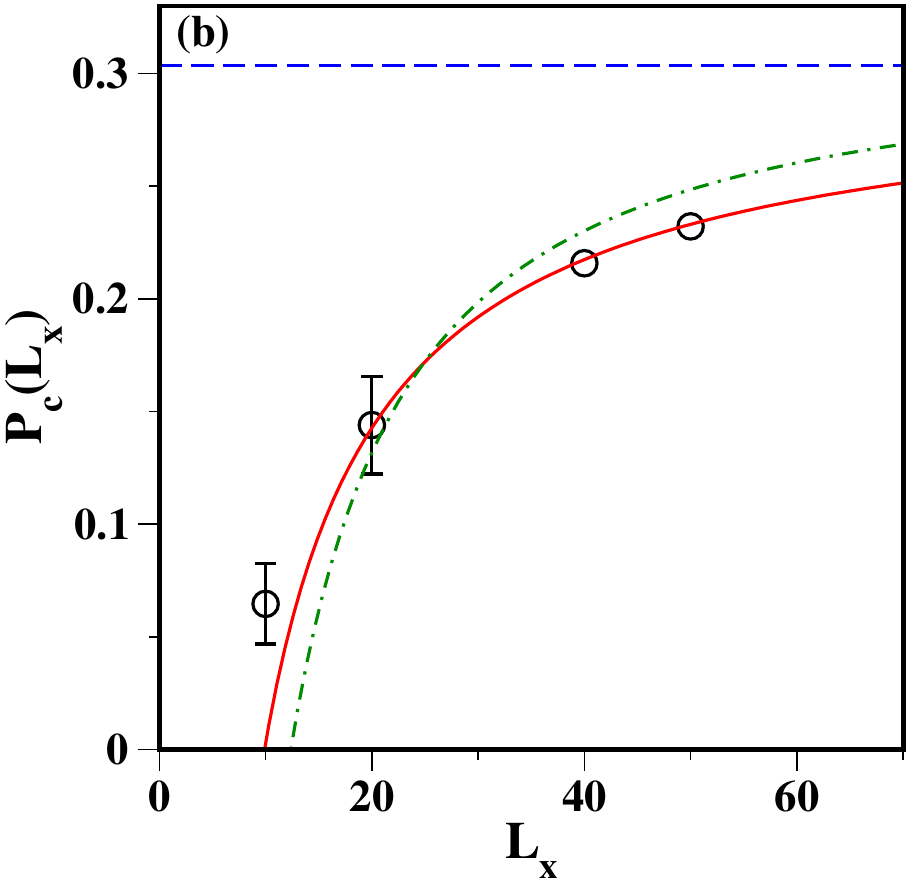}
\includegraphics[width=4.6cm]{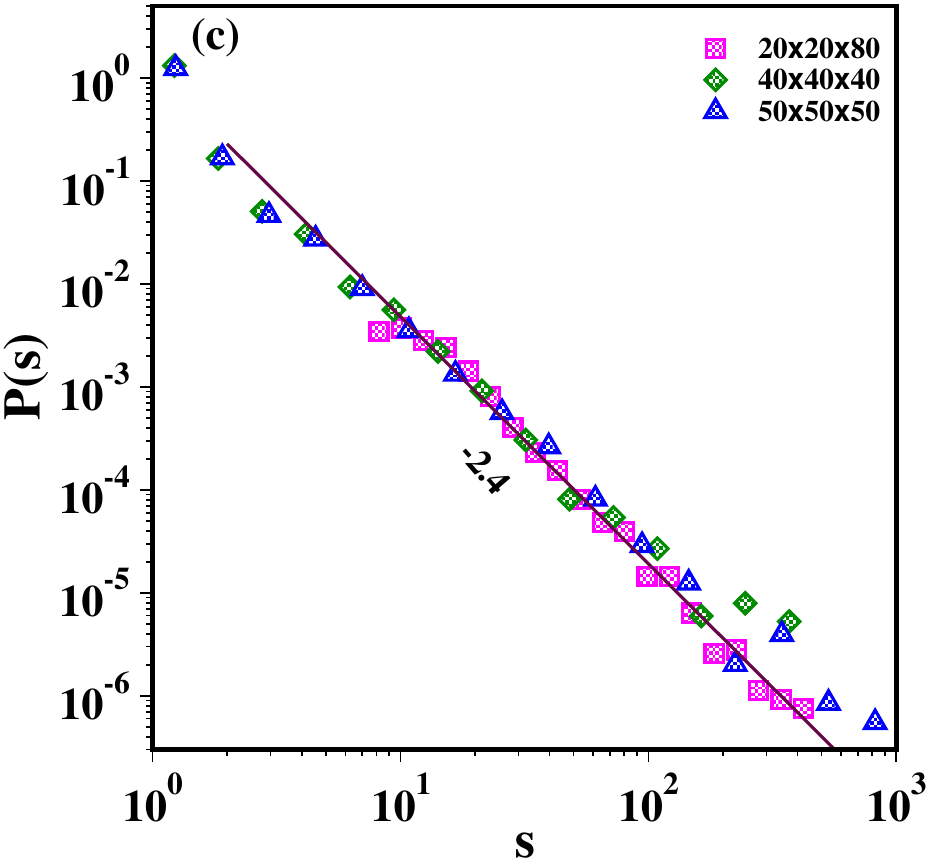}
}
\caption{\label{fig4} Occurrence of directed percolation (DP) transition. (a)
Variation of $p_{\textrm{span}}$ with $p$ for four different system sizes
$10\times10\times40$ (black circles), $20\times20\times80$ (magenta
squares), $40\times40\times40$ (green diamonds), $50\times50\times50$
(blue triangles) for $\dot{\gamma} = 10^{-4}$. (b) Finite size
scaling of critical points for the systems sizes shown in (a), the
red solid line shows the fit according to the scaling law $p_{c}(L) =
0.30339539 -2.412 L^{-1/1.106}$ corresponding to DP. Green dash-dotted
line shows the same scaling function with parameters corresponding to
standard percolation: critical point $p_{c}(\infty) = 3.116$ and $\nu =
0.8765$. The blue dotted line shows the asymptotic $p_{c}(\infty)$ for
DP which is $0.30339539$. (c) Cluster size distribution around $p_{c}$
for the system sizes shown in (a), data shows a power-law decay with
exponent $-2.4$ which is consistent with the prediction for DP.}
\end{figure*}

%%%%%%%%%%%%%%%

Next, for different imposed $\dot{\gamma}$, if we monitor the numerical
growth of mobile cells with increasing strain, a distinct variation
is revealed; see Fig.~\ref{fig3}(c). For example, at a strain of
$\dot{\gamma}t= 0.5$ (marked by orange dotted line), we see that around
$50\%$ of the sites are mobile at low strain rates while the same is close
to $75\%$ at high strain rates.  This implies that subsequent to the
percolation transition (at the critical strain corresponding to $p_c$,
which is around $0.0685$ for $\dot{\gamma}=10^{-4}$), the spatial
heterogeneity of activity is more long-lived for smaller strain rates,
as is shown in Fig.~\ref{fig3}(c), via cuts in the $x-z$ plane of the
local MSD maps corresponding to the evolving trajectories of the states
shown in Fig.~\ref{fig3}(b). While for the largest shear-rate, mobile
regions proliferate in the system, for the smaller shear-rate it is
more localised and takes the form of a well-structured shear band. Thus,
one can infer that the local dynamics, post-percolation, changes with
decreasing shear-rate.

%%%%%%%%%%%%%%%%%%%

To clarify the nature of the percolation transition, we determine
the critical point for the percolation process, $p_c$, using finite
size scaling.  In Fig.~\ref{fig4}(a), we show how $p_{\rm span}$
varies with $p$ for four different system sizes - we observe that the
onset of percolation shifts to larger values of $p$ with increasing
system size.  By fitting the obtained threshold for different system
sizes, using the finite size scaling function $p_{c}(L) = p_{c}(\infty) +
bL^{-1/\nu_{\parallel}}$ (see Fig.~\ref{fig4}(b), red line), we obtain an
estimate of $p_{c}(\infty) = 0.30339539$ and the exponent $\nu_{\parallel}
= 1.106$, which corresponds to a DP transition \cite{wd13}. To compare,
the corresponding numbers for standard percolation are $p_{c}(\infty)$
($= 0.3116$) and $\nu$ ($= 0.8765$); the finite-size scaling function
using these parameters does not fit our data (see Fig.~\ref{fig4}(b),
green line).  Furthermore, we compute the size distribution of clusters
of mobile cells in the vicinity of the percolation transition
(Fig.~\ref{fig4}(c)). As expected, the distribution has a power-law
shape, with the corresponding DP exponent of $2.39$ well characterising
the distribution \cite{carvalho88,mz99}.  Thus, the percolation of the
active regions, in this regime of flow, is a directed process, driven
in the direction of the external shear.

%%%%%%%%

%%%%%%%%%%

\begin{figure*}
\centerline{
\begin{tabular}{cc}
\includegraphics[width=4.5cm]{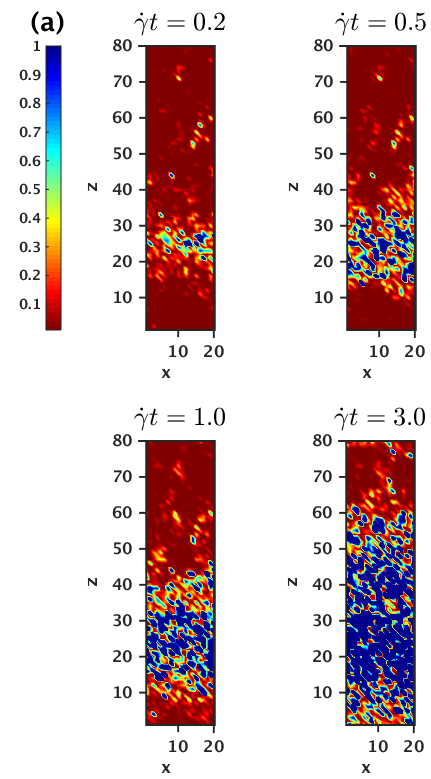}&
\includegraphics[width=8.0cm]{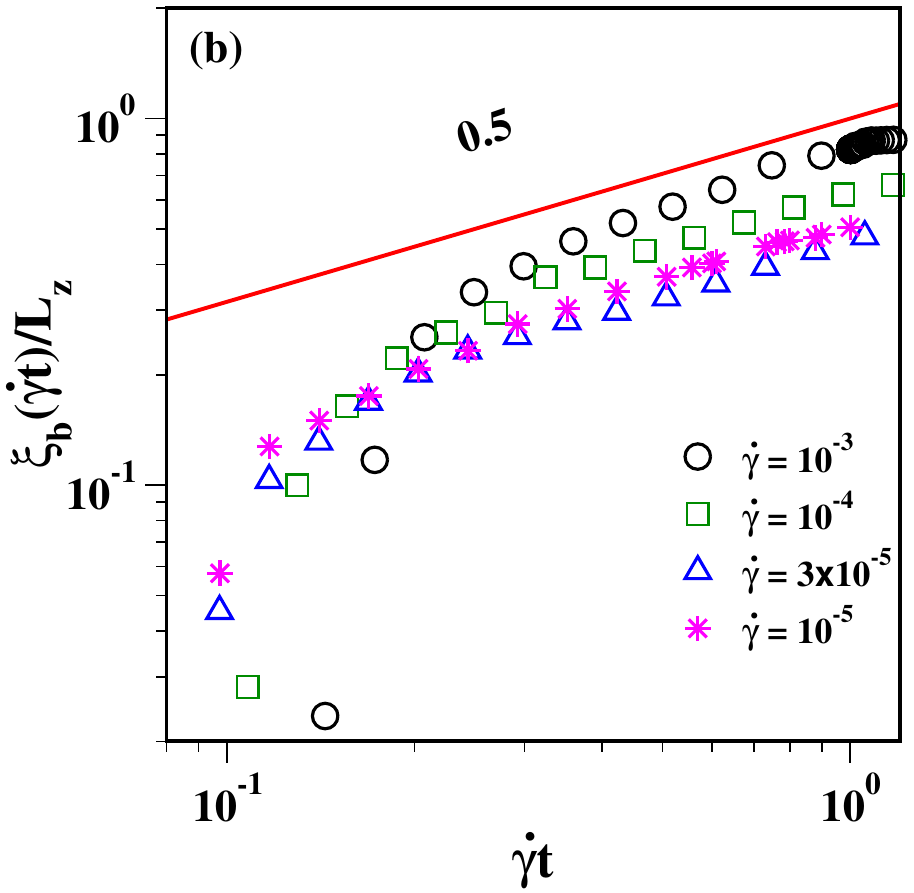}\\
\end{tabular}
}
\caption{\label{fig5} (a) Two dimensional projections of MSD maps showing
the evolution of shear band with time, for imposed $\dot{\gamma}=10^{-4}$,
at strains of $\dot\gamma{t}=0.2, 0.5, 1.0, 3.0$.  (b) Variation of width
of the band $\xi_{b}$, scaled by the box length $L_z$, with $\dot{\gamma}t
=10^{-3}, 10^{-4}, ~3\times10^{-5}~\textrm{and} ~10^{-5}$. The solid
red line corresponds to $t^{1/2}$.}
\end{figure*}

%%%%%%%%%%%

Thus far, we have discussed the existence of an underlying percolation
process of the mobile regions, which occurs for the entire range of
$\dot{\gamma}$ that we have studied. Now, we will focus on how the
dynamics proceeds once the percolating cluster has formed.  In order
to quantify that, we construct two-dimensional projections of the MSD
maps. In Fig.~\ref{fig5}(a), we show the time evolution of the local
mobility, via such maps, for $\dot{\gamma}=10^{-4}$.  We observe that the
$z$-width of the shear band increases with time and eventually the entire
system becomes mobilised. By locating contiguous layers of mobility, we
identify the shear band and, thereafter, by marking the interfaces of
this band, we measure how the band-width, $\xi_{\textrm{b}}$, evolves
with time. For different imposed shear rates, this time evolution is
shown in Fig.~\ref{fig5}(b). We see that $\xi_{\textrm{b}}$ initially
grows quickly and then eventually it reaches a regime where the data
can be fitted with $\xi_{\textrm{b}} \sim t^{1/2}$, implying that
the propagating interface of the shear band has a diffusive motion.
The diffusion constant is dependent on the imposed shear rate. The
smaller the shear rate $\dot{\gamma}$, the slower is the diffusion, which
leads to more long-lived heterogeneities, as discussed earlier. For
the largest shear rate shown, $\dot{\gamma}=10^{-3}$, the diffusive
regime is very short-lived as the band quickly spans the entire system.
For even larger shear rates, beyond percolation, mobile regions quickly
appear everywhere and fluidizes the whole system, and, as a consequence,
a shear band is not clearly discernible.

%%%%%%%%%%%%%%%%%%%%%%%%%%%%%%

%%%%%%%%%%%%%%%%%%%%%

{\it Conclusions.} In this work, we have explored how a model glass,
subject to a constant strain rate, evolves from a quiescent state
to plastic flow. We have shown that this process is initiated by a
DP transition. Under shear, hot spots form in the amorphous solid,
i.e.~local regions in the system where particles have undergone large
non-affine displacements.  These are local structural changes transforming
the initial quiescent glassy state which thus cannot be regained via
thermal fluctuations. Unlike liquids, this is an essential feature of the
glassy state, where a strong non-linear response to the external shear
over-rides thermal fluctuations.  Under continuing deformation, such hot
spots percolate in the direction of the applied drive. Such a scenario
conforms to the DP conjecture \cite{janssen81,grassberger82}, whereby a
system, with no quenched disorders, transforms from a fluctuating state
into an absorbing state. In our case, the dominance of the external
shear leads to the system getting irreversibly trapped in an absorbed
state of the percolating cluster of hot spots.

The subsequent growth of the active regions after the DP transition
depends strongly on the applied shear rate. At larger shear
rates, we observe a quick proliferation of mobile spots leading to
a fluidisation. On the other hand, the process is asymptotically much
slower at smaller shear-rates, with the mobile front slowly diffusing
into the rest of the material. Thus, the existence of an initial fast
timescale for the burst in mobility in the flow direction along the
neutral plane and the subsequent slow timescale for the spread in the
shear-gradient direction, leads to the sustenance of shear bands with
long timescales for small shear rates.

\section{Methods}
We consider a binary mixture of Lennard-Jones (LJ) particles (say A and
B) with 80:20 ratio. This is a well-studied glass former. Particles
interact via LJ potential which is defined as:
\begin{eqnarray}
\label{LJ1}
\textrm{U}^{\textrm{LJ}}_{\alpha\beta}(r) &=& 
\phi_{\alpha\beta}(r)-\phi_{\alpha\beta}(R_{c})-\left(r-R_{c}\right)\left. 
\frac{d\phi_{\alpha\beta}}{dr}\right|_{r=R_{c}},\nonumber\\
\phi_{\alpha\beta}(r) &=& 
4\epsilon_{\alpha\beta}\left[\left(\sigma_{\alpha\beta}/r\right)^{12}-
\left(\sigma_{\alpha\beta}/r\right)^{6}\right]\: r<R_{c},
\end{eqnarray}
where $\alpha, \beta = \textrm{A, B}$. Interaction among
particles is defined as $\epsilon_{\textrm{AA}} = 1.0$,
$\epsilon_{\textrm{AB}} = 1.5\epsilon_{\textrm{AA}}$,
$\epsilon_{\textrm{BB}} = 0.5\epsilon_{\textrm{AA}}$. Range
of interactions is given as $\sigma_{\textrm{AA}} = 1.0$,
$\sigma_{\textrm{AB}} = 0.8\sigma_{\textrm{AA}}$, $\sigma_{\textrm{BB}} =
0.88\sigma_{\textrm{AA}}$ and $R_{c} = 2.5\sigma_{\textrm{AA}}$. Masses
of both type of particles are equal, i.e., $m_{\textrm{A}}
= m_{\textrm{B}} = m$. All quantities are expressed in LJ units
in which the unit of length is $\sigma_{\textrm{AA}}$, energy is
expressed in the units of $\epsilon_{\textrm{AA}}$ and the unit of time
is $\sqrt{{m\sigma_{\textrm{AA}}^{2}}/\epsilon_{\textrm{AA}}}$. More
details about the model and parameters can be found in Ref.~\cite{ka94}.

We perform molecular dynamics (MD) simulation in the $NVT$ ensemble
using the package LAMMPS (``Large-scale Atomic/Molecular Massively
Parallel Simulator'') \cite{plimpton95}.  Different geometries
are considered, placing the particles in boxes of dimensions
$10\times10\times40$, $20\times20\times80$, $40\times40\times40$ and
$50\times50\times50$. Temperature is maintained via dissipative particle
dynamics (DPD) thermostat \cite{sk03}.

Our method for the preparation of glass is as follows: First we
equilibrate high temperature initial configuration at a temperature $T=0.45$
in the super-cooled regime and then quench it to a temperature $T = 0.2$
below the mode coupling transition temperature \cite{ka94}. We wait until
$t_{\textrm w} = 10^4$ and apply shear on $x$-$z$ plane in the direction
of $x$ with different constant strain rates $\dot{\gamma} = 10^{-2},
10^{-3}, 3\times10^{-4}, 10^{-4}, 3\times10^{-5}$ and $10^{-5}$. To
simulate a sheared bulk glass, we use Lees-Edwards periodic boundary
conditions \cite{le72}.

{\bf Maps of local MSD.} To get the MSD maps, we divide the simulation
box into small cubic sub-boxes having linear size of $\sigma_{\textrm{AA}}$. 
At any time $t$, we calculate the average MSD of  the particles populating each sub-box
at $t=0$ (unsheared glassy state). As discussed earlier,  we measure the 
$z$-component of MSD of each particle. The corresponding three-dimensional map,
thus constructed, is shown in the different plots.

{\bf Maps of local strain.} To plot the strain maps, again we divide
the simulation box into small cubic sub-boxes, as in the case of MSD maps. The
numerical derivative of $z$-component of displacement with respect to
$x$, i.e. $\epsilon_{zx}=\partial{\Delta{r_z}}/\partial{x}$ is calculated. This derivative, which gives strain
$\epsilon_{zx}$, is plotted for each sub-box to construct the map.

{\bf Identifying the interface of shear-band.} To characterize shear
bands, we choose a slightly higher threshold on MSD. In the present work,
it is taken as $\mu_{\rm th} = 0.1$. We divide the simulation box into $x$-$y$
layers of thickness one particle diameter and calculate $\mu$, which is
the $z$-component of MSD, for each layer. We assign each layer a value
$\psi=$1/0 depending on whether $\mu$ is greater/lower than $\mu_{th}$. 
To get the size of shear band we count the number of adjoining
layers for which $\psi = 1$.

\section{Acknowledgement}
We acknowledge financial support by the Deutsche Forschungsgemeinschaft (DFG)
in the framework of the priority programme SPP 1594 (Grant No. HO 2231/8-1).

\section{Author Contributions}
All authors contributed to designing and performing the simulations, analysing the data
and writing the manuscript.


\begin{thebibliography}{99}
\bibitem{note} 
There is a slight dependence of results on the choice of threshold but
the qualitative behaviour does not change.
\end{thebibliography}
\end{document}